\documentclass[aps,prapplied,reprint,amsmath,amssymb,superscriptaddress,floatfix]{revtex4-1}
\usepackage{graphicx}
\usepackage[colorlinks, citecolor=blue, linkcolor=blue]{hyperref}

\newcommand{\CNN}{Centre de Nanosciences et de Nanotechnologies, CNRS, Universit\'e Paris-Saclay, 91120 Palaiseau, France}
\newcommand{\IPCMS}{Institut de Physique et Chimie des Mat\'eriaux de Strasbourg, CNRS, Universit\'e de Strasbourg, 67034 Strasbourg, France}

\begin{document}

%
%
\title{Quantifying the thermal stability in perpendicularly magnetized ferromagnetic nanodisks with forward flux sampling}

\author{L. Desplat}
\email{louise.desplat@ipcms.unistra.fr}
\affiliation{\CNN}
\affiliation{\IPCMS}
\author{J.-V. Kim}
\email{joo-von.kim@c2n.upsaclay.fr}
\affiliation{\CNN}

\date{\today}

%
%
\begin{abstract}
The thermal stability in nanostructured magnetic systems is an important issue for applications in information storage. From a theoretical and simulation perspective, an accurate prediction of thermally-activated transitions is a challenging problem because desired retention times are on the order of 10 years, while the characteristic time scale for precessional magnetization dynamics is of the order of nanoseconds. Here, we present a theoretical study of the thermal stability of magnetic elements in the form of perpendicularly-magnetized ferromagnetic disks using the forward flux sampling method, which is useful for simulating rare events. We demonstrate how rates of thermally-activated switching between the two uniformly-magnetized ``up'' and ``down'' states, which occurs through domain wall nucleation and propagation, vary with the interfacial Dzyaloshinskii-Moriya interaction, which affect the energy barrier separating these states. Moreover, we find that the average lifetimes differ by several orders of magnitude from estimates based on the commonly assumed value of 1 GHz for the attempt frequency.
\end{abstract}

\maketitle

%
%
\section{Introduction}
The thermal stability of magnetic states is a challenging problem that underpins the utility of magnetism for information storage. It involves understanding the average retention (or dwell) time of a magnetic bit, which typically comprises regions of uniform magnetization in hard disk media or uniform states in magnetoresistive random access memories. This retention time, $\tau$, can be obtained as the inverse of the transition rate, $\tau^{-1} = k$, which is governed by an Arrhenius relation of the form \cite{Hanggi:1990en}
\begin{equation}
k = f_0 \exp\left( -\frac{\Delta E}{k_B T} \right),
\end{equation}
where $f_0$ is the Arrhenius prefactor often referred to as the ``attempt frequency'', $\Delta E$ is the energy barrier separating the binary `0' (``up'') and `1' (``down'') states, $k_B$ is Boltzmann's constant, and $T$ is the temperature. The vast majority of studies to date have focused on understanding and optimizing the barrier $\Delta E$, whilst assuming a nominal value of $f_0 \simeq$ 1 GHz which captures the typical time scales of damped precessional magnetization in strong ferromagnets. Based on this assumption, the typical metric of a 10-year retention time demands that $\Delta E / k_B T \simeq 50$ for operation at room temperature, a rule-of-thumb that has provided guidance both theoretically and experimentally for the feasibility of using various magnetic states for information storage~\cite{weller1999thermal, chen2010advances, lederman1994measurement,Sampaio:2016cz,Cortes2017:thermal,Gastaldo:2019jx}.

One example of current interest in which the issues of thermal stability are intertwined with the complexity of nonuniform magnetic states concerns ferromagnetic nanostructures with large perpendicular magnetic anisotropy (PMA). Such systems are attractive because they offer larger storage densities than in-plane magnetized systems~\cite{Bhatti:2017be}. However, the reversal process in such systems can be nonuniform~\cite{Khvalkovskiy:2013ea, ChavesOFlynn:2015de, Munira:2015ka,  Jang:2015cl, Sampaio:2016cz, Devolder:2016ji, Devolder:2016dg, Lavanant:2019gk, Gastaldo:2019jx, Volvach:2020ch}.  For nm-thick ferromagnetic films with lateral dimensions in the tens to hundreds of nm, magnetization reversal takes place through the nucleation and propagation of magnetic domain walls, whose energies then govern the energy barrier required to transition from one metastable state to the other~\cite{Khvalkovskiy:2013ea, ChavesOFlynn:2015de, Munira:2015ka}. Moreover, PMA often involves coupling to strong spin-orbit materials, which can also induce an antisymmetric exchange in the form of a Dzyaloshinskii-Moriya interaction (DMI) \cite{Heinze2011:spontaneous,Moreau2016:additive}. Indeed, the presence of DMI has been shown to be detrimental to the thermal stability in such structures~\cite{Jang:2015cl, Sampaio:2016cz, Gastaldo:2019jx}, which results from the fact that the DMI reduces the domain wall energy~\cite{Heide:2008da} and therefore the energy barrier.

Recent studies on magnetic skyrmions, however, have shown that the barrier alone does not govern their thermal stability~\cite{Wild:2017hj, Bessarab:2018cv, Desplat:2018cg, Desplat:2019dn, vonMalottki:2019eb, Desplat:2020fa}. It has been found that a strong entropic contribution to the prefactor can arise as a result of 
the complex topology of the energy surface resulting from the large number of degrees of freedom present. This can be understood from a generalization of Kramers' transition rate theory~\cite{Kramers:1940kg, Hanggi:1990en} to multidimensional phase spaces, as developed by Langer~\cite{Langer:1969jc}, in which the activation entropy is given by the spectrum of small fluctuations about the initial (meta)stable state and the transition state.
Since the DMI also has a strong influence on the spin wave spectrum in the nominally uniformly-magnetized~\cite{GarciaSanchez:2014dw} and domain wall states~\cite{GarciaSanchez:2015kc, Borys:2015ba}, the degree to which entropic contributions influence the thermal stability of PMA memory elements with domain-wall mediated reversal remains to be explored.

Here, we revisit the question of thermal stability in PMA disks using the method of forward flux sampling (FFS)~\cite{Allen:2005dn,Allen:2006simulating, Allen:2009kb, Borrero:2009dt, Vogler:2013bq, Vogler2015:calculating,Desplat:2020fa}. This method was developed for tackling the problem of rare events, which are unlikely to appear during the course of a conventional simulation run, due to the fact that the mean waiting time between events is much larger than the timescale of the dynamics. In magnetism, FFS has been applied to a similar problem of thermal stability in graded media, where its efficacy with respect to brute-force Langevin simulations was clearly demonstrated \cite{Vogler:2013bq}. The method has also been applied recently to the study of skyrmion lifetimes, where agreement was found with another approach based on Langer's theory~\cite{Desplat:2020fa}.

The remainder of the article is organized as follows. In Section II, we present the system studied and the implementation of the forward flux sampling method. Section III discusses the application of the method to determine the change in average dwell time with the Dzyaloshinskii-Moriya interaction. A discussion and concluding remarks are given in Section IV.

\section{Geometry and method}
The system studied comprises a perpendicularly-magnetized ferromagnetic disk, which simulates the free magnetic layer in a magnetic memory device. Following Ref.~\onlinecite{Sampaio:2016cz}, we consider a CoFeB film with a saturation magnetization of $M_s = 1.03$ MA/m, an exchange constant of $A = 10$ pJ/m, and a perpendicular anisotropy constant of $K_u$ = 0.77 MJ/m$^3$. We also take into account the presence of an interfacial Dzyaloshinskii-Moriya interaction, $D$, whose strength is varied. We model a 1-nm thick disk with a diameter of 32 nm, which is discretized with $32 \times 32 \times 1$ finite difference cells. Let $z$ denote the axis perpendicular to the film, which is defined by the $xy$-plane.

We use the micromagnetic approximation in which we consider the Langevin dynamics of the magnetization vector, $\mathbf{m}(\mathbf{r},t) \equiv \mathbf{M}(\mathbf{r},t)/M_s$, such that $\|\mathbf{m}(\mathbf{r},t)\|=1$. The Langevin dynamics is obtained by stochastic time integration of the Landau-Lifshitz equation with a fluctuating thermal field
\begin{equation}
\frac{d \mathbf{m}}{dt} = -\gamma_0 \mathbf{m} \times \left( \mathbf{H}_\mathrm{eff} + \mathbf{h}_\mathrm{th}  \right) + \alpha \mathbf{m} \times \frac{d \mathbf{m}}{dt},
\label{eq:stocLLG}
\end{equation}
where $\gamma$ is the gyromagnetic constant, $\gamma_0 = \mu_0 \gamma$, and $\alpha$ is the Gilbert damping constant. The dynamics is governed by the deterministic effective field, $\mathbf{H}_\mathrm{eff} = -(\mu_0 M_s)^{-1}\delta U/\delta \mathbf{m}$, which is obtained from the variational derivative of the total micromagnetic energy, $U$, with respect to the magnetization, and a stochastic field, $\mathbf{h}_\mathrm{th}$, which takes into account finite temperature effects. This thermal field has zero mean, $\langle  \mathbf{h}_\mathrm{th} \rangle = 0$, and represents a Gaussian white noise with the spectral properties
\begin{equation}
\langle  h_{\mathrm{th},i}(\mathbf{r},t) h_{\mathrm{th},j}(\mathbf{r}',t') \rangle = \frac{2 \alpha k_B T}{\mu_0 V} \delta_{ij} \delta(\mathbf{r}-\mathbf{r}') \delta(t-t'),
\end{equation}
where $i,j$ represent the different Cartesian components of the field vector. We used two methods to compute this stochastic time integration: a homemade code that employs a Heun scheme~\cite{GarciaPalacios:1998kd,Desplat:2019dn,Desplat:2020fa}, and the MuMax3 code~\cite{Vansteenkiste:2014et} with an adaptive time step scheme~\cite{Leliaert:2017ci}.

Let `A' denote the initial magnetic state, which comprises a uniformly-magnetized state along the $+z$ direction, and `B' the second degenerate metastable state, which is uniformly magnetized along $-z$ and separated from `A' by an energy barrier $\Delta E$.  Our goal is to compute the average lifetime $\tau$ of the state `A' at finite temperatures. Estimating $\tau$ directly through brute-force Langevin-dynamics simulations would involve starting with `A' as the initial configuration and integrating Eq.~(\ref{eq:stocLLG}) until the state `B' is reached. This process would then require to be repeated a few hundred times to obtain reasonable statistics. Because of the precessional dynamics, typical time steps for the numerical time integration are in the range of 1 to 100 fs, so it does not appear fruitful to proceed with this program of work where $\tau$ is of the order of years for technologically-relevant systems.

\begin{figure}
\centering\includegraphics[width=8cm]{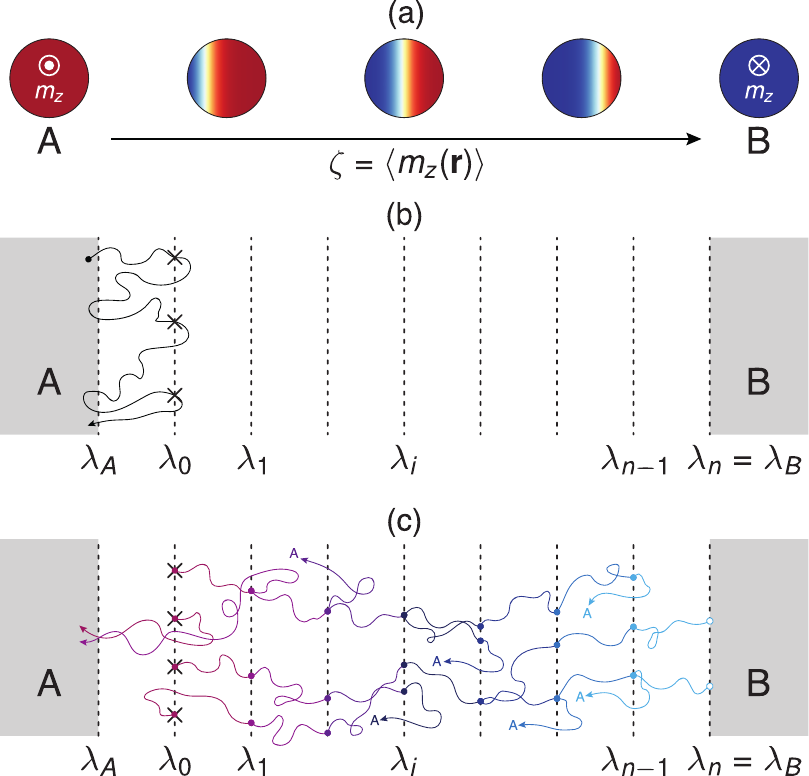}
\caption{Schematic of the forward flux sampling method. (a) Minimum energy path for magnetization reversal between the up and down states through domain wall nucleation and propagation. Progression along the reaction coordinate is parametrized by the order parameter $\zeta$. (b) Sequence of interfaces $\{\lambda \}$ separating the basins `A' and `B'. $N_0$ crossings at the interface $\lambda_0$ are recorded when the system starts in `A' and successfully reaches $\lambda_0$. (c) Simulated trajectories between subsequent interfaces. The micromagnetic configuration is stored at $\lambda_{i+1}$ for each successful crossing from $\lambda_i$ to $\lambda_{i+1}$, which serves as a starting point for Langevin dynamics simulations toward the next interface. Failed crossings involve returns to the basin `A'.}
\label{fig:FFS}
\end{figure}

Path sampling methods such as forward flux sampling (FFS)~\cite{Allen:2005dn, Allen:2006simulating, Allen:2009kb, Borrero:2009dt} can be used instead to estimate the transition rates of rare events such as the thermally-activated escape from `A' to `B'~\cite{Vogler:2013bq, Vogler2015:calculating,Desplat:2020fa}. The basic idea of the FFS approach is illustrated in Fig.~\ref{fig:FFS}. 

The method involves generating trajectories between `A' and `B' in a ratchetlike manner through a series of interfaces $\lambda(\zeta)$ in phase space, $\{ \lambda_\mathrm{A}, \lambda_0, \lambda_1, \ldots, \lambda_{n-1}, \lambda_n = \lambda_\mathrm{B} \}$, which represent non-intersecting isosurfaces of a monotonically varying order parameter $\zeta$. The basin `A' is defined for $\zeta > \zeta_{\mathrm{A}}$, similarly `B' is defined for $\zeta < \zeta_\mathrm{B}$. 
For the problem studied here, the order parameter is chosen to be the spatial average of the $m_z$ component of the magnetization, $\zeta \equiv \langle m_z(\mathbf{r})\rangle = (V)^{-1} \int dV \; m_z(\mathbf{r})$, and decreases monotonically between `A' and `B'. A choice of order parameter that is close to the reaction coordinate -- i.e., the minimum energy path (MEP) of highest statistical weight through phase space -- improves the computational efficiency of the method. All trajectories from `A' to `B' must traverse each interface at least once and the overall transition rate is expressed as
\begin{equation}\label{eq:k_ffs}
k_\mathrm{AB} = \Phi_{\mathrm{A},0} P_B,
\end{equation}
where $\Phi_{\mathrm{A},0}$ represents the flux of trajectories from `A' to the first interface $\lambda_0$, and $P_B \equiv P(\lambda_\mathrm{B} | \lambda_{0})$ is the probability that a trajectory that crossed $\lambda_0$ coming from `A' reaches $\lambda_\mathrm{B}$ before returning to the basin `A'. The quantity $\Phi_{\mathrm{A},0}$ can be obtained in a straightforward manner since the trajectories emanating from `A' cross the $\lambda_0$ interface with high frequency as a result of its proximity. On the other hand, the probability $P_B$ will typically be very small for rare events. However, calculating this probability becomes manageable by decomposing it into a product of partial fluxes at each interface,
\begin{equation}
P_B \equiv P(\lambda_{n} | \lambda_{0}) = \prod_{i=0}^{n-1}{P(\lambda_\mathrm{i+1} | \lambda_{i})},
\end{equation}
where the conditional probability $P(\lambda_\mathrm{i+1} | \lambda_{i})$ represents the probability that a trajectory starting at $\lambda_i$ reaches $\lambda_{i+1}$ before returning to the basin `A'.

Calculation of the FFS rate $k_\mathrm{AB}$ proceeds in two steps. The first involves initiating the system in `A' and performing a Langevin dynamics simulation by stochastic time integration of Eq.~(\ref{eq:stocLLG}). This simulation is used to collect micromagnetic configurations at $\lambda_0$ that result from instances in which the system commences in `A' and crosses the interface $\lambda_0$, as shown in Fig.~\ref{fig:FFS}(b). One then waits for the system to return to `A' before the next crossing configuration at $\lambda_0$ can be saved. The simulation proceeds until $N_0$ such configurations have been obtained, at which the flux of trajectories out of `A' that cross $\lambda_0$ can be estimated as $\Phi_{\mathrm{A},0} = N_0/\Delta t_\mathrm{sim}$, where $\Delta t_\mathrm{sim}$ is the total simulated time required to obtain the $N_0$ crossings. Note that $\Delta t_\mathrm{sim}$ should not include the time the system may have spent in the `B' basin. The second step involves computing the probability $P_B$. This begins by selecting  at random  a stored configuration at $\lambda_0$ and performing a Langevin dynamics simulation until either the system returns to `A', which counts as a failed crossing, or it reaches the next interface $\lambda_1$, in which case the micromagnetic configuration at $\lambda_1$ is stored. This process is repeated $M_0$ times, and the probability of reaching $\lambda_1$ from $\lambda_0$ is computed as $P(\lambda_1 | \lambda_0) = N_0^s/M_0$, where $N_0^s$ denotes the number of successful crossings of $\lambda_1$. The procedure then continues in an analogous manner for the subsequent interfaces until $\lambda_\mathrm{B}$ is reached, as shown in Fig.~\ref{fig:FFS}(c). The overall FFS simulation is successful if at least one trajectory reached $\lambda_\mathrm{B}$.

\section{Variation of thermal stability with DMI}


\subsection{Underdamped limit with full dipolar interactions}\label{sec:underdamped_DDI}
We first present results using the MuMax3 code with full dipolar interactions, a realistic damping value of $\alpha=0.01$~\cite{Bilzer:2006ke}, a variable DMI constant $D$ between 0 and 2 mJ/m$^2$, and $T = 300$ K. For this set of simulations, we used $N_0 = 50$ and $M_0 = 1000$ with 16 interfaces ($n = 15$). The interfaces were constructed as follows. The boundary of the basin `A', $\lambda_\mathrm{A}$, was determined by the median of the value of $m_{z,0} = \langle m_z \rangle$ of the thermally equilibrated state, which leads to `A' being defined as $\zeta > m_{z,0}$. By symmetry, we define $\lambda_\mathrm{B}$ at $\zeta_B = -m_{z,0}$, with `B' occupying the region of phase space $\zeta < -m_{z,0}$. We note that $m_{z,0}$ is a temperature-dependent quantity and also varies strongly as a function of the DMI, since the boundary conditions at the disk edges result in a canting of the magnetic moments away from the $z$ axis~\cite{Rohart:2013ef, GarciaSanchez:2014dw}.

Instead of spacing the interfaces equally from $\zeta = m_{z,0}$ to $\zeta = -m_{z,0}$, we chose instead a weighting function based on the $\tanh(\zeta)$ function that better mimics the reversal path. 
For example, for a magnetization profile of the form  $\langle m_z\rangle (\chi) = \tanh(\chi)$, where $\chi$ is the reaction coordinate \cite{Desplat:2020}, the choice of interface would represent regular spacing along the $\chi$ axis, rather than $\langle  m_z \rangle $. Ultimately, this is a matter of convenience as the FFS method is not overly reliant on the particular placement of interfaces, which should in principle only affect the efficiency of the method \cite{Allen:2009kb}, as we show further below.

\begin{figure}
\centering\includegraphics[width=7cm]{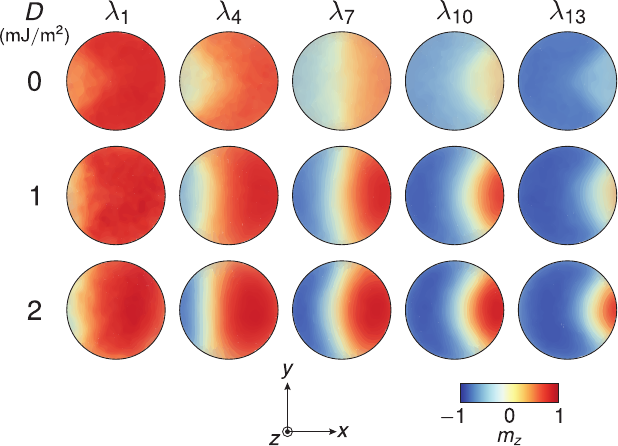}
\caption{Averaged magnetization configurations at different interfaces $\lambda$ for different runs involving three values of the DMI, $D$. $\lambda_7$ corresponds to the interface at which $\zeta = 0$. Each configuration is rotated about the disk center such that wall propagation proceeds from left to right during reversal before the averaging procedure is performed.}
\label{fig:mint}
\end{figure}

In Fig.~\ref{fig:mint}, we present the averaged magnetization configurations at different interfaces for three values of $D$.
The configurations represent the stored state at an interface $\lambda_i$ following a successful traversal from $\lambda_{i-1}$. Since wall nucleation is not restricted to a particular edge of the disk, there are a multitude of paths in which the wall can traverse the disk during the reversal. In order to obtain a meaningful average, we first rotate each configuration such that wall displacement takes place from left to right during the reversal, in line with the schematic presented in Fig.~\ref{fig:FFS}(a). The overall behavior is similar for all values of $D$ considered and corresponds to the minimum energy path predicted in earlier work~\cite{ChavesOFlynn:2015de, Sampaio:2016cz}. The first interface involves the apparition of a nucleation zone at the disk boundary, which is most pronounced for $D = 2$ mJ/m$^2$ by virtue of the greater tilt of the magnetization at the disk edges. This proceeds with the propagation of a domain wall that sweeps through the disk, which is also symmetric about the $\lambda_7$ interface at which $\zeta = 0$. The domain wall exhibits a stronger curvature at the disk center for the largest value of the DMI considered, because the DMI lowers the energy of the curved wall with respect to a straight wall, and selects a preferred direction of the curvature, in the same way that it selects right- or left-rotating N\'eel skyrmions depending on the sign of $D$. The configurations in Fig.~\ref{fig:mint} show that the trajectories generated by Langevin dynamics resemble the minimum energy path obtained with path finding schemes such as the nudged elastic band method \cite{Jonsson1998:nudged,Bessarab:2015method,Sampaio:2016cz,Desplat:2020} or the string method \cite{Weinan2002:string,Jang:2015cl}, which typically constitutes the path of largest statistical weight. We note that in FFS simulations, the thermal fluctuations at 300 K consistently excite the wall's curving modes at the saddle point for large DMI, while in MEPs the relaxed wall in the center tends to remain straight for all values of $D$.

\begin{figure}
\centering\includegraphics[width=8.5cm]{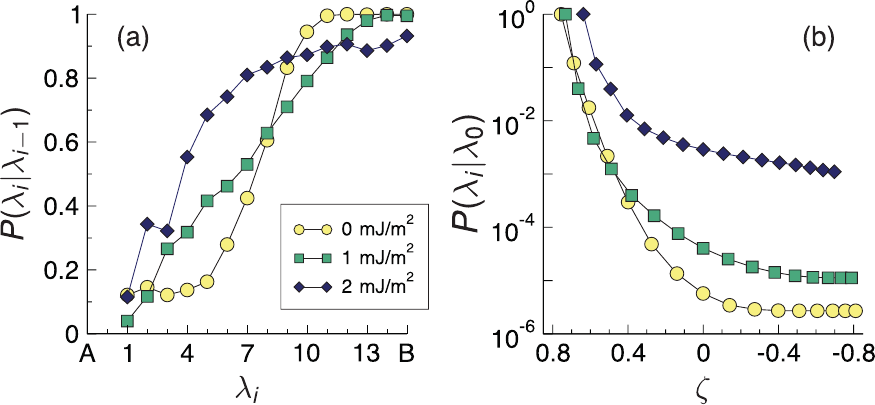}
\caption{(a) Conditional probabilities $P(\lambda_{i} | \lambda_{i-1})$ as a function of the interface $\lambda_{i}$ for three different values of the DMI, $D$. (b) Cumulative conditional probabilities $P(\lambda_i | \lambda_0)$, determined from the cumulative products of $P(\lambda_{i}|\lambda_{i-1})$ in (a), as a function of the order parameter $\zeta$ for three values of $D$.}
\label{fig:pint}
\end{figure}

The properties of the conditional probabilities $p_{i-1} \equiv
 P(\lambda_{i}|\lambda_{i-1})$ are presented in Fig.~\ref{fig:pint}.
The variation of $p_{i-1}$ with the order parameter $\zeta_i$ is shown in Fig.~\ref{fig:pint}(a). 
For the first set of interfaces in the vicinity of `A', the conditional probabilities lie are around 10\%, and then progressively increase as the trajectories propagate toward $\lambda_\mathrm{B}=\lambda(\zeta_{\mathrm{B}})$. For $D = 0$ mJ/m$^2$, there is a sharp increase after $\zeta\simeq 0.4$ $(\lambda_5)$ and saturation is attained after $\zeta\simeq -0.4$ $(\lambda_{11})$. The variation is more gradual for the cases with finite values of $D$, where interestingly for $D = 2$ mJ/m$^2$ saturation toward unity does not occur. This suggests that numerous recrossings of the barrier take place even when the state approaches the vicinity of `B'. This is a general feature of all the cases studied, albeit to different degrees, which is evidenced by the fact that $p_{i-1}$ approaches 1 well beyond the interface where $\zeta = 0$, i.e., at the top of the energy barrier. In other words, the underdamped dynamics of the magnetization precession means that reversal is not guaranteed even if the energy barrier is crossed~\cite{GarciaPalacios:1998kd}. This can also be seen in the evolution of the cumulative product $P(\lambda_i | \lambda_0)$ with $\zeta$, shown in Fig.~\ref{fig:pint}(b), which describes the probability of reaching $\lambda_i$ given the starting point of $\lambda_0$. This function is constructed from the conditional probabilities in Fig.~\ref{fig:pint}(a) as $P(\lambda_i \lvert \lambda_0) = \prod_{j=0}^{i-1} P(\lambda_{j+1} \lvert \lambda_{j})$. In this representation, we observe that most of the evolution in $P(\lambda_i | \lambda_0)$ occurs for $\zeta > 0$, i.e., as the energy barrier is surmounted from `A', with a further reduction by a factor of $\sim$ 5 occurring due to barrier recrossings for $\zeta < 0$.

%
\subsection{Overdamped limit with effective perpendicular anisotropy}\label{sec:overdamped_Keff}
The data presented in Figs.~\ref{fig:mint} and \ref{fig:pint} represent a single FFS run. While the computational time required to execute this task was considerably shorter than full Langevin dynamics simulations of the entire reversal process, it was not feasible to obtain statistics of the interface probabilities, i.e., different realisations of the sampling for a given value of $D$, within a reasonable timeframe. For instance, each data point for the conditional probability in Fig.~\ref{fig:pint}(a) required 1 to 3 days of simulation time on a single NVIDIA GTX 1080 graphics processor unit. Part of this difficulty stems from the calculation of the long-range dipolar interactions, which is computationally intensive. Another difficulty involves the underdamped nature of the precessional dynamics, which results in multiple recrossings at any given interface.

\begin{figure}
\centering\includegraphics[width=8.5cm]{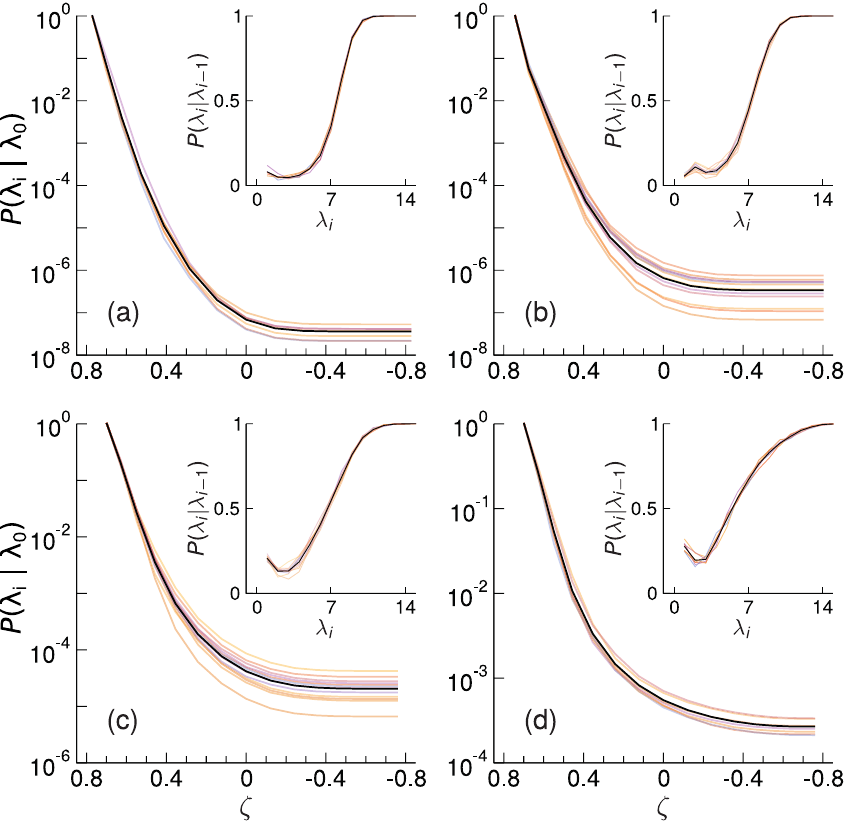}
\caption{Cumulative probability, $P(\lambda_i|\lambda_0)$, as a function of the order parameter, $\zeta$, for different values of $D$: (a) 0.5 mJ/m$^2$, (b) 1.0 mJ/m$^2$, (c) 1.5 mJ/m$^2$, and (d) 2.0 mJ/m$^2$. The insets show the conditional probability $P(\lambda_{i}|\lambda_{i-1})$ as a function of the interface $\lambda_i$. The colored lines represent single FFS runs, while the solid black lines represent averages over the different runs.}
\label{fig:pint_od}
\end{figure}

To ascertain whether the trends observed in Fig.~\ref{fig:pint} are representative of the reversal process, we conducted a different set of simulations in which the long-ranged dipolar interactions are approximated by a rescaling of the perpendicular anisotropy constant, i.e., $K_\mathrm{eff} = K_u - \frac{1}{2}(N_z-N_x)\mu_0 M_s^2$ = 187 kJ/m$^3$, where $N_x = 0.0418$ and $N_z = 0.916$ are demagnetizing factors of the disk \cite{Chen1991:demagnetizing}. The use of a local approximation for the dipolar interaction is less computationally intense and speeds up the simulation time. We first considered the overdamped limit, $\alpha = 0.5$, which facilitates comparisons with Langer's theory~\cite{Desplat:2020}. The cumulative probability function $P(\lambda_i|\lambda_0)$ for different values of the DMI is shown in Fig.~\ref{fig:pint_od}.

The corresponding conditional probabilities $P(\lambda_{i}|\lambda_{i-1})$ are shown in the inset of each subfigure. For each value of $D$, between 7 and 12 distinct FFS runs were performed and are represented by the different colored curves, while the ensemble-averaged result is given by the black curve.  While there is some spread in the final probability $P(\lambda_B|\lambda_0)$, the curves for the cumulative probability and the conditional probability share the same qualitative trend for a given value of $D$. The manner in which the shape of the $P(\lambda_{i}|\lambda_{i-1})$ curves change as $D$ is increased also mirrors the evolution seen for the case presented in Fig.~\ref{fig:pint}(b), where the `S'-shaped variation at $D=0$ transitions to more of a square-root like behavior at $D = 2$ mJ/m$^2$.

\subsection{Details of the optimization in the underdamped limit with effective perpendicular anisotropy}

\paragraph{Optimization}
In the underdamped limit of $\alpha=0.01$, similar to Section~\ref{sec:underdamped_DDI}, recrossings are more frequent and the system deviates more significantly from the MEP, resulting in longer waiting times between magnetization reversals compared to the overdamped case. Longer waiting times typically imply longer CPU times. Nevertheless, the FFS simulation times can be reduced through several methods. 

First, many trial runs at a given interface can be performed simultaneously on one or multiple CPU units. The only requirement is that the simulation waits for all trials run at one interface to terminate before moving on to the next interface. This can be managed via MPI on multiple nodes, or a simple Bash script on a single node. 

\begin{figure}
\centering\includegraphics[width=8.5cm]{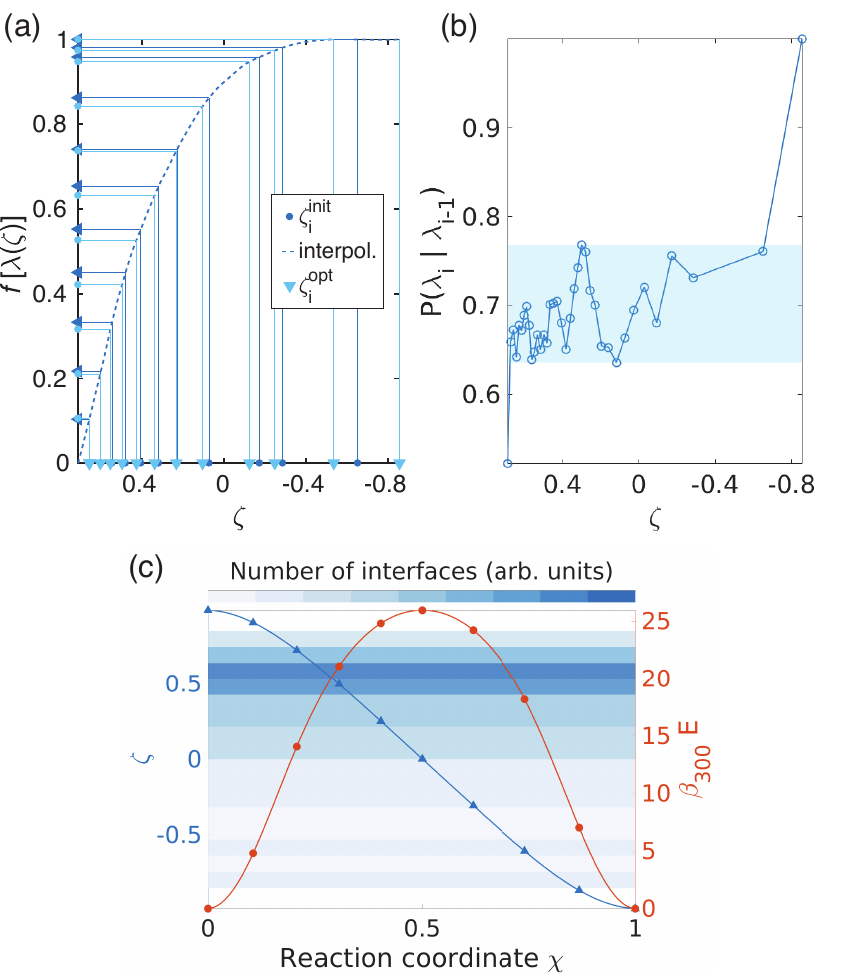}
\caption{ Example of interface optimization in FFS.  (a) The new values of the order parameter, $\{\zeta_i^{\mathrm{opt}}\}$, are obtained by inverting the interpolation function in Eq.~(\ref{eq:ffs_interpol_function}) which is plotted in a dotted line. The  $\{\zeta_i^{\mathrm{opt}}\}$ are designed to approach a constant flux of partial trajectories through each interface. (b) Partial fluxes $P(\lambda_i \lvert \lambda_{i-1})$ as a function of the order parameter $\zeta$ resulting from the interface optimization procedure. With the exception of the first and last interfaces, the values are found within the shaded area.  (c) Evolution of the order parameter $\zeta$ (left), and of the Energy normalized by the thermal energy at 300 K, $\beta_{300} E$ (right), as a function of the reaction coordinate $\chi$ along the minimum energy path. The interfaces are optimized again after the first FFS run with the interfaces shown in (a). The number of interfaces per interval of $\zeta$ is represented by the colored background, where a darker color corresponds to more interfaces.}
\label{fig:opt_and_pi}
\end{figure}

Second, the parameters of the FFS run can be optimized in order to improve the efficiency of the method, as derived by Borrero and Escobedo \cite{Borrero:2009dt}.  The variance of the rate  yielded by the trial runs, i.e., $V[P_B]$ \cite{Allen:2006simulating}, is minimized under the constraint that the total rate $k_{AB}$ must remain constant. The best efficiency for FFS is obtained for a minimal relative variance of the rate, which is yielded by a constant flux of partial trajectories through all interfaces, i.e., $M_i p_i = \mathrm{cst} $. We define the relative variance of the rate as~\cite{Allen:2006simulating},
\begin{equation}
\mathcal{V} = \sum_{i=0}^{n-1} \frac{q_i}{p_i k_i},
\label{eq:variance}
\end{equation}
where we have used the simplified notations $p_i \equiv  P(\lambda_{i+1}| \lambda_i)$, $q_i \equiv  1-p_i$, and $k_i \equiv M_i/N_0$. One may therefore either optimize the number of trial runs per interface, $\{M_i\}$, at fixed interface placement, or optimize the placement of the interfaces, $\{\zeta_i\}$, at fixed $\{M_i\}$. In the latter case, the interface placement is used to optimize the conditional probabilities, $\{p_i\}$. In this work, we chose to set $M_i=M_0$ at all interfaces, and optimize the interface placement. We proceeded as follows. First, a full FFS simulation was carried out with $N_0=100$ and $n+1=40$ interfaces whose positions in phase space were chosen as described in Section~\ref{sec:underdamped_DDI}. Note that once $\Phi_{A,0}$ in Eq.~(\ref{eq:k_ffs}) has been determined, then if an estimate of the final rate $k$ is available -- for example from the Kramers' method \cite{Langer:1969jc, Desplat:2020, Desplat:2020fa}, or from previous runs with similar parameters -- it is then possible to  estimate a value of $M_0$ that yields a low relative variance of the rate $\mathcal{V}$, as defined in Eq.~(\ref{eq:variance}). For instance, one may aim for $\mathcal{V}=1$. For the present set of parameters, we used values of $M_0$ spanning from 1000 at high $D$, to 2800 at low $D$. Once the initial run has terminated, the interface placement is optimized to obtain $p_i = p_i^{\mathrm{opt}} = (P_B)^{\frac{1}{n}}$ $\forall  i$, by following the procedure described in~\cite{Borrero:2009dt}, and illustrated in Fig.~ \ref{fig:opt_and_pi}.
A monotonic interpolation function is needed in order to establish a one-to-one correspondence between the $\{p_i\}$ and the $\{ \zeta_i\}$. A possible choice of such a function is,
\begin{equation}\label{eq:ffs_interpol_function}
f[\lambda(\zeta_i)]= \frac{\sum_{j=0}^{i-1} \ln p_j}{\sum_{j=0}^{n-1} \ln p_j},
\end{equation}
which reduces to the optimal values $f( \zeta_i^{\mathrm{opt}} )  = i/n$ when $p_i =p_i^{\mathrm{opt}}$. The initial set of $f( \zeta_i^{\mathrm{init}} )$ is computed from the $\{ \zeta_i^{\mathrm{init}} \}$ according to Eq.~(\ref{eq:ffs_interpol_function}). The sets of $ \zeta_i^{\mathrm{init}} $ and $f( \zeta_i^{\mathrm{init}} )$ are respectively shown as dark blue dots and dark blue triangles in  Fig.~ \ref{fig:opt_and_pi}a.
The resulting data set is interpolated, where the interpolation function is shown as a dotted line.
One then computes the optimal values, $f(\lambda_i^{\mathrm{opt}})$, as represented by pale blue dots. The corresponding values of the order parameter $\{\zeta_i^{\mathrm{opt}}\}$ that determine the placement of the new interfaces are found by inverting the interpolation function, and are shown as pale blue triangles. Note that,  for readability, we do not show all the interfaces in Fig.~ \ref{fig:opt_and_pi}a. In practise, only a single FFS run was typically carried out per value of $D$, and the computed probabilities were then used to optimize the interfaces for the next value of $D$, and so on with iteratively decreasing $D$.

Fig.~\ref{fig:opt_and_pi}b shows the partial flux through interface $i$ coming from interface $i-1$, $ p_{i-1}$, as a function of $\zeta$ at $D$ = 0.75 mJ/m$^2$. The interface placement is the one obtained from the interface optimization procedure carried out using the results at  $D$ = 1 mJ/m$^2$. Apart from the flux through $\lambda_1$ and $\lambda_n=\lambda_B$, we have $0.64 \leq p_{i-1} \leq 0.77$, as shown by the shaded area in Fig.~\ref{fig:opt_and_pi}b. For $N_0=100$, $M_0=2800$, the relative variance of the rate yielded by the FFS run is $\mathcal{V}=1.37$ and the computed lifetime of the uniform state at 300 K is $\tau = 0.013$ $\pm 0.002$ s.

Lastly, Fig.~\ref{fig:opt_and_pi}c shows the evolution of the order parameter, $\zeta$, and of the energy normalized by the thermal energy at $T=300$ K, $\beta_{300} E$, in which $\beta_{300}=(k_BT_{300})^{-1}$ is Boltzmann's factor at 300 K, as a function of the normalized reaction coordinate, $\chi$, as computed by the geodesic nudged elastic band  method (GNEB)~\cite{Bessarab:2015method} in Ref.~\onlinecite{Desplat:2020}. $\chi$ goes from 0 in state `A', to 0.5 at the barrier top, to 1 in state `B'. The colored background indicates the number of interfaces per small interval of $\zeta$ with arbitrary units, with a darker color corresponding to more interfaces. During the initial FFS run, the interfaces were equally spaced out along the reaction coordinate, similarly to the images in the GNEB method, which are shown as blue triangles. The optimized interfaces are not evenly spaced out but are denser about halfway to the saddle point ($\zeta \simeq 0.58$ and  $\chi \simeq 0.27$). In terms of the energy profile, this corresponds to the region preceding the barrier top. The fact that more interfaces are needed in that area is likely caused by the sudden variation in the flux of partial trajectories, similar to the ones shown in Fig.~\ref{fig:pint}. 

\begin{figure}
\includegraphics[width=.7\linewidth]{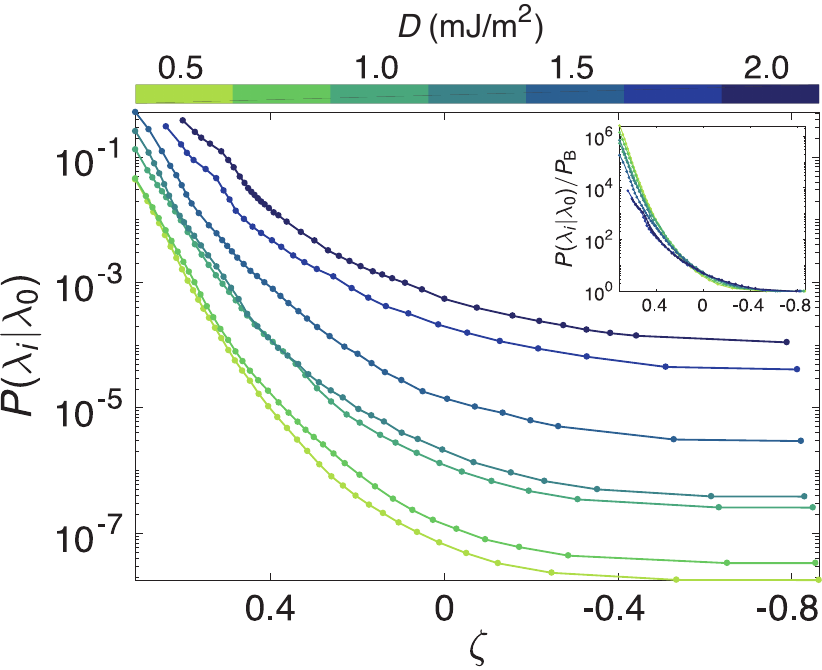}
\caption{Cumulative probability, $P(\lambda_i | \lambda_0)$, as a function of the order parameter, $\zeta$, for different values of D. The inset shows the same probabilities normalized by $P_B = P(\lambda_B \lvert \lambda_0)$. The normalized probabilities for different values of $D$ are superimposed because the flux of barrier re-crossings remains a constant fraction of the rate constant, independently of $D$.
\label{fig:cumul_fluxes_alpha_d0d01}}
\end{figure}

\paragraph{Results}
The Langevin simulations in the underdamped regime were performed with a homemade code~\cite{Desplat:2019dn, Desplat:2020fa} with an effective anisotropy and a single FFS run per value of $D$. Each run was carried out on a single CPU unit (Intel Xeon Gold 6130 processor), and a maximum of 65 simultaneous Langevin trial runs. A complete run at $T$ = 300 K and $n=39$ interfaces took between about a week at high $D$ ($M_0 \simeq 1000$), to about two weeks for low $D$ ($M_0 \simeq 2800$).  In Fig. \ref{fig:cumul_fluxes_alpha_d0d01}, we show, for a set of trajectories starting at $\lambda_0$, the cumulative probability to reach $\lambda_i$,  $P(\lambda_i \lvert \lambda_0)$, as a function of the order parameter $\zeta$, for different values of $D$. 
We find that overall, the behavior of the system is the same as in the overdamped case of Section~\ref{sec:overdamped_Keff}, with profiles of the cumulative fluxes similar to the ones shown in Fig. \ref{fig:pint_od}. Once more, the global flux of trajectories decreases with decreasing $D$.   The inset in Fig.~\ref{fig:cumul_fluxes_alpha_d0d01} shows the cumulative fluxes normalized by $P_B$, i.e.,  $P(\lambda_i \lvert \lambda_0)/P_B$. We find that in this new viewpoint, all graphs are superimposed.  Since the energy profile is symmetric, the rate of recrossing remains a constant fraction of the transition rate, and the DMI does not fundamentally impact the system in that sense.

%
\subsection{Comparison of lifetimes}
\begin{figure}
\centering\includegraphics[width=8.5cm]{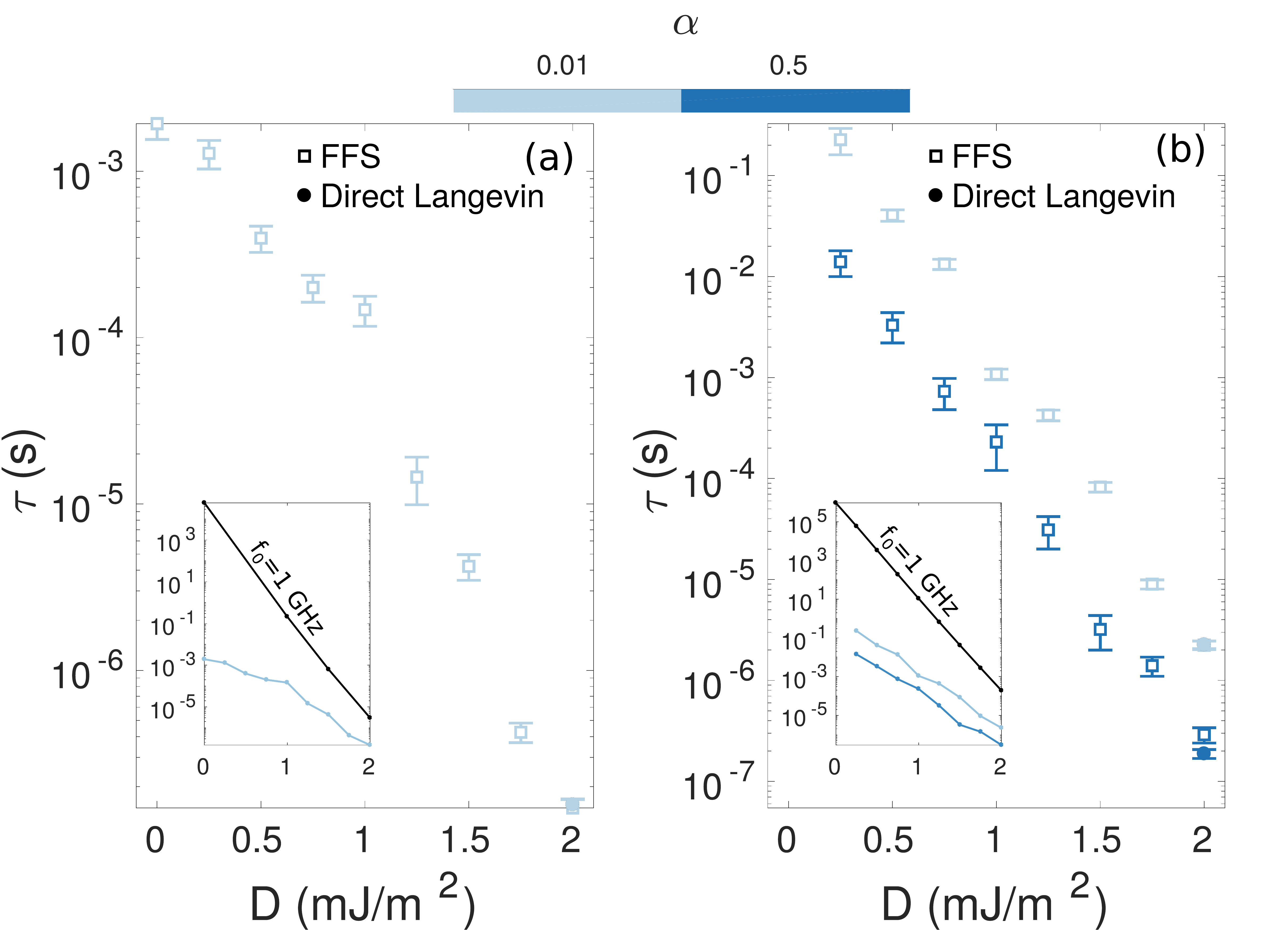}
\caption{Average lifetime, $\tau$, as a function of the DMI, $D$, at $T = 300$ K computed with forward flux sampling (FFS) and direct Langevin dynamics simulations. (a)  Lifetimes for $\alpha = 0.01$ with full dipole-dipole interactions. (b) Lifetimes for $\alpha = 0.01$ and $\alpha = 0.5$ with effective perpendicular anisotropy. The average lifetime computed from full Langevin dynamics simulations is given for all three cases at $D = 2$ mJ/m$^2$. The insets give a comparison of the FFS data with the constant $f_0 = 1$ GHz approximation with the energy barriers respectively computed in Refs.~\onlinecite{Sampaio:2016cz} and \onlinecite{Desplat:2020}.}
\label{fig:tauvarD}
\end{figure}

%
The key result of this study is presented in Fig.~\ref{fig:tauvarD}, where the mean lifetime of the `A' state, $\tau$, is shown as a function of $D$ for the three cases discussed above. For memory elements, it corresponds to the information retention time. 
In Fig.~\ref{fig:tauvarD}(a), the results for the simulations with full dipolar interactions are shown. Since these results were obtained from only a single FFS run, we estimate the uncertainty in the lifetime by using the relative variance of the rate $\mathcal{V}$, as defined in Eq.~(\ref{eq:variance}). We recall that $M_i = M_0 = 1000$ for all interfaces for this case and $N_0 = 50$. The statistical error in the rate is given by~\cite{Allen:2006simulating},
\begin{equation}\label{eq:k_err}
    \sigma_k=k_{AB} \sqrt{\frac{\mathcal{V}}{N_0}},
\end{equation}
and is thus assumed to arise only from the trial runs. As anticipated from previous work, the lifetime appears to decrease with increasing DMI, which can be attributed to the linear decrease in the energy barrier with $D$ that is directly related to the variation in the domain wall energy (per unit surface area), $\sigma_w = 4\sqrt{A K_u} - \pi D$. The striking result here however is the magnitude of $\tau$, which varies from a few milliseconds at $D = 0$ to a few tenths of a microsecond at $D = 2$ mJ/m$^2$. These values are in stark contrast with the $\tau$ predicted by taking $f_0 = 1$ GHz with the energy barriers computed elsewhere~\cite{Sampaio:2016cz}, as shown in the inset of Fig.~\ref{fig:tauvarD}(a). This discrepancy stems from a strong entropic contribution to the prefactor~\cite{Desplat:2020}, which results in a slower decrease of the overall lifetime as the barrier is reduced. In the present example, the constant $f_0$ approximation predicts a decrease by 10 orders of magnitude in $\tau$ as $D$ is varied from $0$ to 2 mJ/m$^2$, whereas the FFS results indicate a change of only 4 orders of magnitude over the same range. As we have shown elsewhere~\cite{Desplat:2020}, compensation effects underpin this behavior and the present example serves to highlight the importance of quantifying the Arrhenius prefactor in magnetic nanostructures.

Fig.~\ref{fig:tauvarD}(b) presents the lifetimes calculated with the effective anisotropy approximation for the underdamped ($\alpha = 0.01$) and overdamped ($\alpha = 0.5$) limits. In the underdamped case, we observe that the overall lifetimes are two orders of magnitude greater than those shown in Fig.~\ref{fig:tauvarD}(a). This discrepancy is primarily due to the difference in energy barriers when the full dipolar interactions are taken into account, which amounts to a reduction of the barrier by about 4 $k_B T$ when dipolar interactions are included~\cite{Sampaio:2016cz, Desplat:2020}. As for the Arrhenius prefactor, we have found that, under a full treatment of the DDI, it is lower than in the effective aniostropy treatment for large $D$, while it is larger for low $D$ \cite{Desplat:2020}. This leads to the apparent two regimes that can be seen in Fig. \ref{fig:tauvarD}(a).
As in Fig.~\ref{fig:tauvarD}(a), the uncertainties in the lifetime were determined from Eq.~(\ref{eq:k_err}) for $\alpha=0.5$, whereas for $\alpha=0.01$, they were obtained from several distinct FFS runs. The latter method yields a larger error, because Eq.~(\ref{eq:k_err}) neglects the statistical error that may stem from an incomplete sampling in the `A' basin, and treats the trial runs as the only source of error.
In the overdamped limit, the lifetimes decrease with the same overall slope as in the underdamped case, but are an order of magnitude shorter. This results from fewer barrier recrossings as compared with the underdamped case, where the domain wall exhibits more of a unidirectional motion as it sweeps through the disk. The system also relaxes faster, so the overdamped trajectories should, on average, remain closer to the MEP.

\begin{figure}
\centering\includegraphics[width=8.5cm]{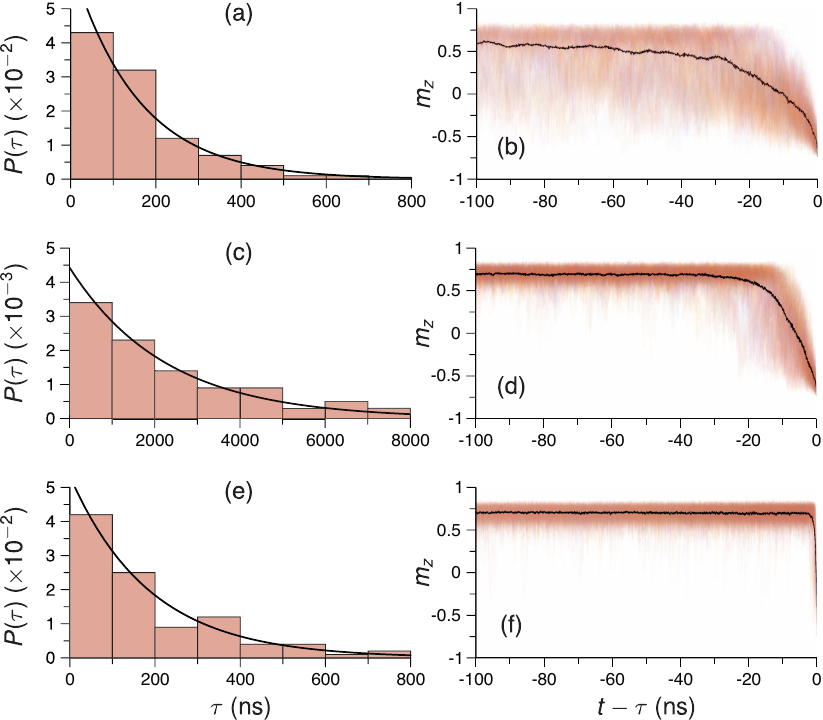}
\caption{Direct Langevin dynamics simulations of the dwell time, $\tau$, for $D = 2$ mJ/m$^2$. The results correspond to (a,b) $\alpha = 0.01$ with full dipolar interactions, (c,d) $\alpha = 0.01$ with effective perpendicular anisotropy, and (e,f) $\alpha = 0.5$ with effective perpendicular anisotropy. For each case, 100 simulations were performed with a successful switching event. (a,c,e) Probability distribution of the dwell time. The solid line represents a fit to an exponential function. (b,d,f) $m_z$ component of the magnetization as a function of time, where the curves are shifted along the time axis such that the first crossing of $-m_{z,0}$ occurs at $t = 0$. The colored curves represent the individual Langevin dynamics simulations, while the solid black line represents an average of these curves.}
\label{fig:langevin}
\end{figure}

A comparison to direct Langevin simulations is possible for the lowest barriers studied, i.e., at $D = 2$ mJ/m$^2$, where the overall lifetimes remain accessible to direct integration of the Eq.~(\ref{eq:stocLLG}) for the full thermally-activated switching process. These data are also presented in Fig.~\ref{fig:tauvarD}, where good quantitative agreement with the FFS results can be seen. The statistics and averaged time traces from these direct Langevin simulations are presented in Fig.~\ref{fig:langevin}.
For each case studied, we performed 100 simulations of successful switching events with mean lifetimes in the range of 10 ns to 220 $\mu$s. The mean values are determined from exponential fits to the probability density distributions of the lifetime shown in Figs.~\ref{fig:langevin}(a), \ref{fig:langevin}(c), and \ref{fig:langevin}(e), while the uncertainties are determined from the variance. The time series data of $m_z$, shown for up to 100 ns before the full switching event, are presented in Figs.~\ref{fig:langevin}(b), \ref{fig:langevin}(d), and \ref{fig:langevin}(f). The underdamped limit is characterized by large fluctuations in the order parameter, where excursions to $m_z < 0$ can be seen frequently well before full reversal occurs. This behavior is most pronounced when full dipolar interactions are included [Fig.~\ref{fig:langevin}(b)] and is indicative of recrossing processes that ultimately lead to a longer average dwell times in the `A' state. 
In contrast, the overdamped limit exhibits fewer excursions of this nature, where the time window in which large fluctuations occur are more localized to the switching event. This can also be seen in the solid black lines superimposed on the time series data, which represent averages over the 100 instances of the simulations and provide a measure of the reproducibility of the switching event. The averages for the underdamped cases exhibit a gradual decrease in $m_z$ toward switching, while the variation is significantly sharper in the overdamped case.

\begin{figure}
\centering\includegraphics[width=8.5cm]{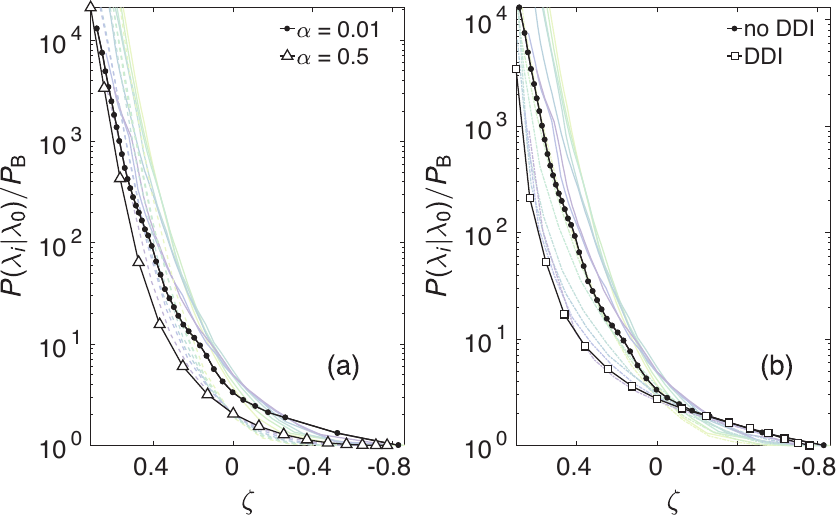}
\caption{Effects of (a) damping and (b) dipole-dipole interactions (DDI) on the cumulative forward fluxes normalized by $P_B$. The transparent colored lines show the cumulative fluxes for all values of $D$ as a function of the order parameter, where the color of the line represents the value of $D$ with the same colorscale as that of Fig.~\ref{fig:cumul_fluxes_alpha_d0d01}. In both graphs, the solid colored lines correspond to the underdamped case with effective anisotropy, while the dashed colored lines corresponds to the other case. The black lines show the average cumulative flux as a function of the average order parameter for each case. }
\label{fig:mean_cumul_fluxes}
\end{figure}

In light of the previous observations, we show in black solid lines in Fig.~\ref{fig:mean_cumul_fluxes},  the cumulative fluxes of trajectories in FFS simulations normalized by $P_B$, and averaged over all values of the DMI, as a function of the average order parameter, $\zeta$.
The transparent colored lines correspond to the individual normalized cumulative fluxes for each value of $D$, where the color of the line corresponds to the value of the DMI according to the colorscale in Fig.~\ref{fig:cumul_fluxes_alpha_d0d01}. Fig.~\ref{fig:mean_cumul_fluxes}(a) corresponds to the fluxes obtained with the effective anisotropy treatment of the dipole-dipole interactions, with two values of the Gilbert damping, respectively $\alpha=0.01$ (underdamped regime) and  $\alpha=0.5$ (overdamped regime). As one can expect, the magnetization reverses more efficiently in the overdamped regime compared to the underdamped case, with an overall larger flux of forward trajectories along the order parameter and a faster saturation past the saddle point at $\zeta<0$, which indicates fewer barrier recrossings. This is in line with the behavior of the magnetization in direct Langevin simulations presented in Figs.~\ref{fig:langevin}(d) and ~\ref{fig:langevin}(f).
Fig.~\ref{fig:mean_cumul_fluxes}(b) shows the averaged cumulative fluxes of trajectories at a constant damping of $\alpha=0.01$, with the effective anisotropy, and the full treatment of dipole-dipole interactions. In the portion of configuration space situated before the saddle point ($\zeta>0$), the effect of DDI on the average fluxes resembles that of the large Gilbert damping in Fig.~\ref{fig:mean_cumul_fluxes}(a), with larger forward fluxes than in the effective anisotropy treatment. Nevertheless, the Langevin dynamics of the system under DDI, as shown in Fig.~\ref{fig:langevin}(b), is drastically different from that of Fig.~\ref{fig:langevin}(f). Since DDI result in large fluctuations in the order parameter in the region of the `A' basin, with $0.8 \gtrsim m_z \gtrsim -0.4$, crossings of the interfaces close to `A', and up to the barrier top, are much more frequent. It follows that the portion of trajectories crossing subsequent interfaces in the $\zeta>0$ region, as opposed to the ones returning to `A', is larger when DDI are included. This explains the larger fluxes in Fig.~\ref{fig:mean_cumul_fluxes}(b). Past the barrier ($\zeta<0$), the ratios of crossing trajectories over the ones returning to `A' appear more similar with and without DDI, with similar amplitudes of the fluctuations during the reversal process in Figs.~\ref{fig:langevin}(b) and \ref{fig:langevin}(d), and the forward fluxes become comparable.


\section{Discussion and Concluding Remarks}


In this work, we have presented an example of how rate constants for magnetic systems may be obtained by means of forward flux sampling simulations. The case detailed here concerns the mean waiting times between magnetization reversals in a nanodisk with parameters similar to a free CoFeB layer, as used in magnetic random access memories, as a function of increasing DMI. In the large DMI case with the largest rate constant, FFS accurately reproduced the result of brute-force direct Langevin simulations, with and without a full treatment of dipole-dipole couplings. When full DDI are taken into account, and under a realistic Gilbert damping of $\alpha=0.01$, the lifetimes of the uniform states decrease with increasing DMI, and span from a  few  milliseconds to tenths of a microsecond, while the stability factor varies from $\Delta=32$ to $\Delta=8$ \cite{Sampaio:2016cz}. In all the cases studied above, we find that the assumption $f_0 = 1$ GHz for the Arrhenius prefactor~\cite{Gastaldo:2019jx,Sampaio:2016cz}
is not justified, and does not hold \cite{Desplat:2020}. 

In this particular class of ultrathin films, for which magnetization reversals take place via the nucleation and propagation of a domain wall, we have found that the average transition path obtained from forward flux sampling simulations closely reproduces the minimum energy path computed by path finding methods. In more complex systems, e.g., thicker films where DDI modify the transition paths, many paths from one stable state to the other may be statistically relevant \cite{Gastaldo:2019jx,Weinan2003:Energy}. Additionally, the path involving the lowest internal energy barrier may not correspond to the dominant transition mechanism, because the activation entropy of the transition accounts for a significant portion of the associated rate constant, and must therefore be taken into account~\cite{Desplat:2018cg,Desplat:2020}. In such systems, FFS should prove particularly useful, as it directly samples the configuration space to compute rate constants, and the paths of highest statistical weight will naturally also be the ones most followed during the trial runs. Additionally, FFS can be used to obtain stationary distributions as a function of the order parameter, and thus observables allowing for comparison with experiments~\cite{Allen:2009kb}. FFS is most efficient when the choice of the order parameter is able to closely mimic the reaction coordinate. In the present work, we chose the average $z$-component of the magnetization, $\zeta = \langle m_z \rangle $, which relates to the reaction coordinate $\chi$ as $\zeta = \tanh \chi$. While this choice of $\zeta$ has proved successful, it is possible that $\zeta = \tanh \langle m_z \rangle$ may be an even better choice in this case.


\section*{Acknowledgements}
The authors thank Nicolas Reyren for stimulating discussions and for sharing a dataset for the inset of Fig.~\ref{fig:tauvarD}. This work was supported by the Agence Nationale de la Recherche under Contract No. ANR-17-CE24-0025 (TOPSKY) and the University of Strasbourg Institute for Advanced Study (USIAS) for a Fellowship, within the French national programme ``Investment for the Future'' (IdEx-Unistra).


\bibliography{articles}

\end{document}